\documentclass[preprint,prd,aps,showpacs,showkeys,nofootinbib]{revtex4}
\usepackage{graphicx}
\begin{document}

%\preprint{hep-ph/0412147}

\title{$B_{s,d}^0\rightarrow l^+l^-$ in the minimal gauged $(B-L)$ supersymmetry}

\author{Tai-Fu Feng\footnote{email:fengtf@hbu.edu.cn},
Yu-Li Yan\footnote{email:yychanghe@sina.com.cn},
Hai-Bin Zhang\footnote{email:hbzhang@hbu.edu.cn},
Shu-Min Zhao\footnote{email:zhaosm@hbu.edu.cn}}

\affiliation{Department of Physics, Hebei University, Baoding, 071002, China\\
State Key Laboratory of Theoretical Physics (KLTP),\\
Institute of theoretical Physics, Chinese Academy of Sciences, Beijing, 100190, China}

\begin{abstract}
Complete expressions of effective Hamilton for $b\rightarrow sl^+l^-\;(l=\mu,\;\tau)$ are derived in the framework
of minimal supersymmetric extension of the standard model with local $B-L$ gauge symmetry.
With some assumptions on parameters of the model, a numerical analysis of the supersymmetric
contributions to the branching ratios of $B_s^0\rightarrow l^+l^-\;(l=\mu,\;\tau)$ is presented.
\end{abstract}


\maketitle

\section{Introduction\label{sec1}}
\indent\indent
The study on rare B decays can detect new physics beyond the standard model (SM) since the
theoretical evaluations on relevant physical quantities are not
seriously affected by the uncertainties due to unperturbative QCD  effects.
The LHCb collaboration reports the observed branching ratios of $B_{s,d}^0\rightarrow\mu^+\mu^-$
as \cite{LHCb-Bsmuons}
\begin{eqnarray}
%%%%%%%%%%%%%%%%%%%%%%%%%%%%%%%%%%%%%%%%%%%%%%%%%%%
&&BR(B_{s}^0\rightarrow\mu^+\mu^-)_{_{EXP}}=(3.20^{+1.5}_{-1.2})\times10^{-9}
\;,\nonumber\\
&&BR(B_{d}^0\rightarrow\mu^+\mu^-)_{_{EXP}}<9.4\times10^{-10}\;.
%%%%%%%%%%%%%%%%%%%%%%%%%%%%%%%%%%%%%%%%%%%%%%%%%%%
\label{EXP-BtoSmuons-0}
\end{eqnarray}
Now, Particle Data Group (PDG) gives the observed averages as \cite{PDG}
\begin{eqnarray}
%%%%%%%%%%%%%%%%%%%%%%%%%%%%%%%%%%%%%%%%%%%%%%%%%%%
&&BR(B_{s}^0\rightarrow\mu^+\mu^-)_{_{EXP}}=(3.1\pm0.7)\times10^{-9}
\;,\nonumber\\
&&BR(B_{d}^0\rightarrow\mu^+\mu^-)_{_{EXP}}<6.3\times10^{-10}\;,
%%%%%%%%%%%%%%%%%%%%%%%%%%%%%%%%%%%%%%%%%%%%%%%%%%%
\label{EXP-BtoSmuons}
\end{eqnarray}
where the experimental observable on branching ratio of $B_{s}^0\rightarrow\mu^+\mu^-$
is nicely consistent with the correspondingly SM prediction \cite{Buras0}
\begin{eqnarray}
%%%%%%%%%%%%%%%%%%%%%%%%%%%%%%%%%%%%%%%%%%%%%%%%%%%
&&BR(B_{s}^0\rightarrow\mu^+\mu^-)_{_{SM}}=(3.23\pm0.27)\times10^{-9}\;,
%%%%%%%%%%%%%%%%%%%%%%%%%%%%%%%%%%%%%%%%%%%%%%%%%%%
\label{SM-BtoSmuons}
\end{eqnarray}
and the experimental precision on $B_{d}^0\rightarrow\mu^+\mu^-$
is already of the same order as the SM evaluation
\begin{eqnarray}
%%%%%%%%%%%%%%%%%%%%%%%%%%%%%%%%%%%%%%%%%%%%%%%%%%%
&&BR(B_{d}^0\rightarrow\mu^+\mu^-)_{_{SM}}=(1.07\pm0.10)\times10^{-10}\;.
%%%%%%%%%%%%%%%%%%%%%%%%%%%%%%%%%%%%%%%%%%%%%%%%%%%
\label{SM-BtoDmuons}
\end{eqnarray}
The precise measurements on the rare B-decay processes set more strict
constraints on the new physics beyond SM. The main purpose of investigation of B-decays
is to search for traces of new physics and determine its parameter space.

In all the extensions of
SM, the supersymmetry is considered as one of the most plausible candidates. In the general supersymmetric
extension of SM, new sources of flavor violation may appear in those soft breaking terms \cite{Gabrielli}.
If we believe that the
SM is only an effective theory and the supersymmetry is more fundamental, study on rare B-processes
will definitely enrich our knowledge in this field. But before we can really pin down any new physics effects,
we need to carry out a thorough exploration in this field, not only in SM, but also in supersymmetric models.
Actually the analyses of constraints on parameters in the minimal supersymmetric extensions of the SM (MSSM)
are extensively
discussed in literature. The calculation of the rate of inclusive decay $B\rightarrow X_s\gamma$ is
presented by authors of Refs.~\cite{Ciuchini1,Ciafaloni,Borzumati1} in the two-Higgs doublet model (2HDM).
The supersymmetric effect on $B\rightarrow X_s\gamma$ is discussed in Refs.~\cite{Bertolini1,Barbieri,Borzumati2,Causse,Prelovsek}
and the next-to-leading order (NLO) QCD corrections are given in  Refs.~\cite{Ciuchini2}.
The transition $b\rightarrow s\gamma\gamma$ in the supersymmetric extension of
the standard model is computed in  Ref.~\cite{Bertolini2}. The hadronic B decays \cite{Cottingham} and
CP-violation in those processes \cite{Barenboim} have been discussed also.
The authors of  Ref.~\cite{Hewett} have discussed possibility of observing supersymmetric effects
in rare decays $B\rightarrow X_s\gamma$ and $B\rightarrow X_se^+e^-$ at the B-factory.
Studies on decays $B\rightarrow (K,K^*)\mu^+\mu^-$ in the SM and
supersymmetric model have been carried out in  Refs.~\cite{Ali,Buras1}.
The supersymmetric effects on these processes are very
interesting and studies on them may shed some light on the
general characteristics of the supersymmetric model.
A relevant review can be found in Refs.~\cite{Masiero1,Masiero1-1}. For oscillations of $B_0-\bar{B}_0$ ($K_0-\bar{K}_0$),
calculations have been done in the SM and 2HDM. As for the supersymmetric extension of SM,
the calculation involving the
gluino contributions should be re-studied carefully for gluino has a nonzero mass. At the NLO approximation,
the QCD corrections to the $B_0-\bar{B}_0$ mixing in the supersymmetry model have been discussed also. The
authors of  Refs.~\cite{Ciuchini3,Contion} applied the mass-insertion method to estimate QCD corrections
to the $B_0-\bar{B}_0$ mixing. The calculations including the gluon-mediated QCD were given
in  Ref.~\cite{Krauss}, and later we have re-derived the formulation
by including the contribution of gluinos  \cite{Feng1}.

The discovery of Higgs on the Large Hadron Collider (LHC) implies that we finish the
spectrum of particles predicted by the standard model (SM) now  \cite{CMS,ATLAS}.
One main target of particle physics is testing the SM precisely and searching for
the new physics (NP) beyond it.
Experimentally the LHCb experiment can measure the quantities
of exclusive hadronic, semi-leptonic, and leptonic $B$ and $B_s$ decays
at a high sensitivity \cite{LHCB}.
In addition the measurements on inclusive rare $B$ decay and decays with neutrino
final states will be performed also in two next generation B factories
in near future  \cite{B-factory1,B-factory2}.

In supersymmetry, R-parity is defined through
$R=(-1)^{3(B-L)+2S}$, where $B$, $L$ and $S$ are baryon number,
lepton number and spin respectively for a concerned
field \cite{R-parity,R-parity-1}. In the MSSM with local $U(1)_{B-L}$ symmetry, R-parity is spontaneously
broken when left- and right-handed sneutrinos acquire nonzero
vacuum expectation values (VEVs) \cite{Perez1,Perez2,Perez3,Perez4}.
Meanwhile, the nonzero VEVs of left- and
right-handed sneutrinos induce the mixing
between neutralinos (charginos) and neutrinos (charged leptons). Furthermore,
the MSSM with  local $U(1)_{B-L}$ symmetry naturally predicates two sterile neutrinos
 \cite{Perez6,Senjanovic1,Chang-Feng}, which are favored by the Big-bang
nucleosynthesis (BBN) in cosmology \cite{Hamann}. In other words, there are
exotic sources to mediate flavor changing neutral current processes (FCNC)
in this model.

Here we investigate the FCNC processes with a $B_{s,d}^0\rightarrow l^+l^-\;(l=\mu,\;\tau)$ transition in the
MSSM with local $U(1)_{B-L}$ symmetry, our presentation is organized as follows.
In section \ref{sec2}, we
briefly summarize the main ingredients of the MSSM with
local $U(1)_{B-L}$ symmetry, then present effective Hamilton for  $b\rightarrow sl^+l^-$
in section \ref{sec3} and the decay widths at hadronic scale in section \ref{sec4}, respectively. The numerical analyses are given in section \ref{sec5}, and our
conclusions are summarized in section \ref{sec6}.

\section{The MSSM with local $U(1)_{B-L}$ symmetry\label{sec2}}
\indent\indent
When $U(1)_{B-L}$ is a local gauge symmetry, one can
enlarge the local gauge group of the SM to $SU(3)_{_C}\otimes
SU(2)_{_L}\otimes U(1)_{_Y}\otimes U(1)_{_{(B-L)}}$. In the
model proposed in Refs.~\cite{Perez1,Perez2,Perez3,Perez4}, the
exotic superfields are three generation right-handed neutrinos
$\hat{N}_{_i}^c\sim(1,\;1,\;0,\;1)$. Meanwhile, quantum numbers of
the matter chiral superfields for quarks and leptons are given by
\begin{eqnarray}
&&\hat{Q}_{_I}=\left(\begin{array}{l}\hat{U}_{_I}\\
\hat{D}_{_I}\end{array}\right)\sim(3,\;2,\;{1\over3},\;{1\over3})\;,\;\;
\hat{L}_{_I}=\left(\begin{array}{l}\hat{\nu}_{_I}\\
\hat{E}_{_I}\end{array}\right)\sim(1,\;2,\;-1,\;-1)\;,
\nonumber\\
&&\hat{U}_{_I}^c\sim(3,\;1,\;-{4\over3},\;-{1\over3})\;,\;\;
\hat{D}_{_I}^c\sim(3,\;1,\;{2\over3},\;-{1\over3})\;,\;\;
\hat{E}_{_I}^c\sim(1,\;1,\;2,\;1)\;,
\label{quantum-number1}
\end{eqnarray}
with $I=1,\;2,\;3$ denoting the index of generation. In addition,
the quantum numbers of two Higgs doublets
are assigned as
\begin{eqnarray}
&&\hat{H}_{_u}=\left(\begin{array}{l}\hat{H}_{_u}^+\\
\hat{H}_{_u}^0\end{array}\right)\sim(1,\;2,\;1,\;0)\;,\;\;
\hat{H}_{_d}=\left(\begin{array}{l}\hat{H}_{_d}^0\\
\hat{H}_{_d}^-\end{array}\right)\sim(1,\;2,\;-1,\;0)\;.
\label{quantum-number2}
\end{eqnarray}
The superpotential of the MSSM with local $U(1)_{B-L}$ symmetry is written as
\begin{eqnarray}
&&{\cal W}={\cal W}_{_{MSSM}}+{\cal W}_{_{(B-L)}}^{(1)}\;.
\label{superpotential1}
\end{eqnarray}
Here ${\cal W}_{_{MSSM}}$ is superpotential of the MSSM, and
\begin{eqnarray}
&&{\cal W}_{_{(B-L)}}^{(1)}=\Big(Y_{_N}\Big)_{_{IJ}}\hat{H}_{_u}^Ti\sigma_2\hat{L}_{_I}\hat{N}_{_J}^c\;.
\label{superpotential-BL}
\end{eqnarray}
Correspondingly, the soft breaking terms for the MSSM with local $U(1)_{B-L}$
symmetry are generally given as
\begin{eqnarray}
&&{\cal L}_{_{soft}}={\cal L}_{_{soft}}^{MSSM}+{\cal L}_{_{soft}}^{(1)}\;.
\label{soft-breaking1}
\end{eqnarray}
Here ${\cal L}_{_{soft}}^{MSSM}$ is soft breaking terms of the MSSM, and
\begin{eqnarray}
&&{\cal L}_{_{soft}}^{(1)}=
-(m_{_{\tilde{N}^c}}^2)_{_{IJ}}\tilde{N}_{_I}^{c*}\tilde{N}_{_J}^c
-\Big(m_{_{BL}}\lambda_{_{BL}}\lambda_{_{BL}}+h.c.\Big)
+\Big\{\Big(A_{_N}\Big)_{_{IJ}}H_{_u}^Ti\sigma_2\tilde{L}_{_I}\tilde{N}_{_J}^c+h.c.\Big\}\;,
\label{soft-breaking3}
\end{eqnarray}
with $\lambda_{_{BL}}$ denoting the gaugino of $U(1)_{_{B-L}}$.
After the $SU(2)_L$ doublets $H_{_u},\;H_{_d},\;\tilde{L}_{_I}$ and $SU(2)_L$ singlets $\tilde{N}_{_I}^c$
acquire the nonzero VEVs,
\begin{eqnarray}
&&H_{_u}=\left(\begin{array}{c}H_{_u}^+\\{1\over\sqrt{2}}\Big(\upsilon_{_u}+H_{_u}^0+iP_{_u}\Big)\end{array}\right)\;,
\nonumber\\
&&H_{_d}=\left(\begin{array}{c}{1\over\sqrt{2}}\Big(\upsilon_{_d}+H_{_d}^0+iP_{_d}\Big)\\H_{_d}^-\end{array}\right)\;,
\nonumber\\
&&\tilde{L}_{_I}=\left(\begin{array}{c}{1\over\sqrt{2}}\Big(\upsilon_{_{L_I}}+\tilde{\nu}_{_{L_I}}+iP_{_{L_I}}\Big)\\
\tilde{L}_{_I}^-\end{array}\right)\;,\nonumber\\
&&\tilde{N}_{_I}^c={1\over\sqrt{2}}\Big(\upsilon_{_{N_I}}+\tilde{\nu}_{_{R_I}}+iP_{_{N_I}}\Big)\;,
\label{VEVs}
\end{eqnarray}
the R-parity is broken spontaneously, and the local gauge symmetry $SU(2)_{_L}\otimes U(1)_{_Y}\otimes U(1)_{_{(B-L)}}$
is broken down to the electromagnetic symmetry $U(1)_{_e}$, and the neutral and
charged gauge bosons acquire the nonzero masses as
\begin{eqnarray}
&&m_{_{\rm Z}}^2={1\over4}(g_1^2+g_2^2)\upsilon_{_{\rm EW}}^2\;,\nonumber\\
&&m_{_{\rm W}}^2={1\over4}g_2^2\upsilon_{_{\rm EW}}^2\;,\nonumber\\
&&m_{_{Z_{BL}}}^2=g_{_{BL}}^2\Big(\upsilon_{_N}^2+\upsilon_{_{\rm EW}}^2-\upsilon_{_{\rm SM}}^2\Big)\;.
\label{gauge-masses}
\end{eqnarray}
Where $\upsilon_{_{\rm SM}}^2=\upsilon_{_u}^2+\upsilon_{_d}^2$,
$\upsilon_{_{\rm EW}}^2=\upsilon_{_u}^2+\upsilon_{_d}^2+\sum\limits_{\alpha=1}^3\upsilon_{_{L_\alpha}}^2$,
$\upsilon_{_N}^2=\sum\limits_{\alpha=1}^3\upsilon_{_{N_\alpha}}^2$,
and $g_2,\;g_1$, $g_{_{BL}}$ denote the gauge couplings of
$SU(2)_{_L},\;\;U(1)_{_Y}$ and $U(1)_{_{(B-L)}}$, respectively.

%%%%%%%%%%%%%%%%%%%%%%%%%%%%%%%%%%Begin 1st Revision%%%%%%%%%%%%%%%%%%%%%%%%%%%%%%%%%%%%
To satisfy present electroweak precision observations, we assume
the mass of neutral $U(1)_{_{(B-L)}}$ gauge boson $m_{_{Z_{BL}}}>1\;{\rm TeV}$ which implies
$\upsilon_{_N}>1\;{\rm TeV}$ when $g_{_{BL}}<1$, then we derive $\max((Y_{_{N}})_{ij})\le10^{-6}$
and $\max(\upsilon_{_{L_I}})\le10^{-3}\;{\rm GeV}$ \cite{Perez4}
to explain experimental data on neutrino oscillation.
Considering the minimization conditions at one-loop level,
we formulate the $3\times3$ mass-squared matrix for right-handed sneutrinos as
\begin{eqnarray}
&&m_{_{{\tilde N}^c}}^2\simeq\left(\begin{array}{ccc}
\Lambda_{_{\tilde{N}_1^c}}^2-\Lambda_{_{BL}}^2\;,\;\;&0\;,\;\;
&-{\upsilon_{_{N_1}}\over\upsilon_{_{N_3}}}\Lambda_{_{\tilde{N}_1^c}}^2\\
0\;,\;\;&\Lambda_{_{\tilde{N}_2^c}}^2-\Lambda_{_{BL}}^2\;,\;\;
&-{\upsilon_{_{N_2}}\over\upsilon_{_{N_3}}}\Lambda_{_{\tilde{N}_2^c}}^2\\
-{\upsilon_{_{N_1}}\over\upsilon_{_{N_3}}}\Lambda_{_{\tilde{N}_1^c}}^2\;,\;\;
&-{\upsilon_{_{N_2}}\over\upsilon_{_{N_3}}}\Lambda_{_{\tilde{N}_2^c}}^2\;,\;\;
&{\upsilon_{_{N_1}}^2\Lambda_{_{\tilde{N}_1^c}}^2+\upsilon_{_{N_2}}^2\Lambda_{_{\tilde{N}_2^c}}^2
\over\upsilon_{_{N_3}}^2}-\Lambda_{_{BL}}^2
\end{array}\right)
\label{R-sneutrino-mass}
\end{eqnarray}
with $\Lambda_{_{BL}}^2=m_{_{Z_{BL}}}^2/2+\Delta T_{_{\tilde N}}$.
Where $\Delta T_{_{\tilde N}}$ denotes one-loop radiative corrections
to the right-handed sneutrinos from top, bottom, tau and their supersymmetric
partners  \cite{Chang-Feng}.

\section{Effective Hamilton for  $b\rightarrow sl^+l^-\;(l=\mu,\tau)$\label{sec3}}
\indent\indent
The transition $b\rightarrow s$ is attributed to the effective Hamilton at hadronic scale
\begin{eqnarray}
&&{\cal H}_{_{eff}}=-{4G_{_F}\over\sqrt{2}}V_{_{tb}}V_{_{ts}}^*\Big[C_{_1}{\cal O}_{_1}^c
+C_{_2}{\cal O}_{_2}^c+\sum\limits_{i=3}^6C_{_i}{\cal O}_{_i}
+\sum\limits_{i=7}^{10}\Big(C_{_i}{\cal O}_{_i}+C_{_i}^\prime{\cal O}_{_i}^\prime\Big)
\nonumber\\
&&\hspace{1.5cm}
+\sum\limits_{i=S,P}\Big(C_{_i}{\cal O}_{_i}+C_{_i}^\prime{\cal O}_{_i}^\prime\Big)\Big]\;,
\label{effective-Hamilton}
\end{eqnarray}
where ${\cal O}_{_i},\;(i=1,\;2,\;\cdots,\;10,\;S,\;P)$ and ${\cal O}_{_i}^\prime,\;(i=7,\;8,\;\cdots,\;10,\;S,\;P)$
are defined as \cite{Buras1}
\begin{eqnarray}
&&{\cal O}_{_1}^u=(\bar{s}_{_L}\gamma_\mu T^au_{_L})(\bar{u}_{_L}\gamma^\mu T^ab_{_L})\;,\;\;
{\cal O}_{_2}^u=(\bar{s}_{_L}\gamma_\mu u_{_L})(\bar{u}_{_L}\gamma^\mu b_{_L})\;,
\nonumber\\
&&{\cal O}_{_3}=(\bar{s}_{_L}\gamma_\mu b_{_L})\sum\limits_q(\bar{q}\gamma^\mu q)\;,\;\;
{\cal O}_{_4}=(\bar{s}_{_L}\gamma_\mu T^ab_{_L})\sum\limits_q(\bar{q}\gamma^\mu T^aq)\;,
\nonumber\\
&&{\cal O}_{_5}=(\bar{s}_{_L}\gamma_\mu\gamma_\nu\gamma_\rho b_{_L})\sum\limits_q(\bar{q}\gamma^\mu
\gamma^\nu\gamma^\rho q)\;,\;\;
{\cal O}_{_6}=(\bar{s}_{_L}\gamma_\mu\gamma_\nu\gamma_\rho T^ab_{_L})\sum\limits_q(\bar{q}\gamma^\mu
\gamma^\nu\gamma^\rho T^aq)\;,
\nonumber\\
&&{\cal O}_{_7}={e\over g_{_s}^2}m_{_b}(\bar{s}_{_L}\sigma_{_{\mu\nu}}b_{_R})F^{\mu\nu}\;,\;\;
{\cal O}_{_7}^\prime={e\over g_{_s}^2}m_{_b}(\bar{s}_{_R}\sigma_{_{\mu\nu}}b_{_L})F^{\mu\nu}\;,
\nonumber\\
&&{\cal O}_{_8}={1\over g_{_s}}m_{_b}(\bar{s}_{_L}\sigma_{_{\mu\nu}}T^ab_{_R})G^{a,\mu\nu}\;,\;\;
{\cal O}_{_8}^\prime={1\over g_{_s}}m_{_b}(\bar{s}_{_R}\sigma_{_{\mu\nu}}T^ab_{_L})G^{a,\mu\nu}\;,
\nonumber\\
&&{\cal O}_{_9}={e^2\over g_{_s}^2}(\bar{s}_{_L}\gamma_\mu b_{_L})\bar{l}\gamma^\mu l\;,\;\;
{\cal O}_{_9}^\prime={e^2\over g_{_s}^2}(\bar{s}_{_R}\gamma_\mu b_{_R})\bar{l}\gamma^\mu l\;,
\nonumber\\
&&{\cal O}_{_{10}}={e^2\over g_{_s}^2}(\bar{s}_{_L}\gamma_\mu b_{_L})\bar{l}\gamma^\mu\gamma_5 l\;,\;\;
{\cal O}_{_{10}}^\prime={e^2\over g_{_s}^2}(\bar{s}_{_R}\gamma_\mu b_{_R})\bar{l}\gamma^\mu\gamma_5 l\;,
\nonumber\\
&&{\cal O}_{_S}={e^2\over16\pi^2}m_{_b}(\bar{s}_{_L}b_{_R})\bar{l}l\;,\;\;
{\cal O}_{_S}^\prime={e^2\over16\pi^2}m_{_b}(\bar{s}_{_R}b_{_L})\bar{l}l\;,
\nonumber\\
&&{\cal O}_{_P}={e^2\over16\pi^2}m_{_b}(\bar{s}_{_L}b_{_R})\bar{l}\gamma_5l\;,\;\;
{\cal O}_{_P}^\prime={e^2\over16\pi^2}m_{_b}(\bar{s}_{_R}b_{_L})\bar{l}\gamma_5l\;.
\label{operators}
\end{eqnarray}

At the electroweak energy scale $\mu_{_{\rm EW}}$, the Wilson coefficients
$C_{_{i,NP}}(\mu_{_{\rm EW}})$
from the new physics beyond SM can be found in
Ref.~\cite{Feng2} and elsewhere.

The Wilson coefficients in Eq.~(\ref{effective-Hamilton}) are calculated at the matching
scale $\mu_{_{\rm EW}}$, then evolved down to hadronic scale $\mu\sim m_{_b}$ by the renormalization
group equations. In order to obtain hadronic matrix elements conveniently, we
define effective coefficients  \cite{Buras1}
\begin{eqnarray}
%%%%%%%%%%%%%%%%%%%%%%%%%%%%%%%%%%%%%%%%%%%%%%%%%%%
&&C_{_7}^{eff}={4\pi\over\alpha_{_s}}C_{_7}-{1\over3}C_{_3}-{4\over9}C_{_4}
-{20\over3}C_{_5}-{80\over9}C_{_6}
\;,\nonumber\\
%%%%%%%%%%%%%%%%%%%%%%%%%%%%%%%%%%%%%%%%%%%%%%%%%%%
&&C_{_8}^{eff}={4\pi\over\alpha_{_s}}C_{_8}+C_{_3}-{1\over6}C_{_4}
+20C_{_5}-{10\over3}C_{_6}
\;,\nonumber\\
%%%%%%%%%%%%%%%%%%%%%%%%%%%%%%%%%%%%%%%%%%%%%%%%%%%
&&C_{_9}^{eff}={4\pi\over\alpha_{_s}}C_{_9}+Y(q^2)
\;,\nonumber\\
%%%%%%%%%%%%%%%%%%%%%%%%%%%%%%%%%%%%%%%%%%%%%%%%%%%
&&C_{_{10}}^{eff}={4\pi\over\alpha_{_s}}C_{_{10}}\;,\;\;\;
C_{_{7,8,9,10}}^{\prime eff}={4\pi\over\alpha_{_s}}C_{_{7,8,9,10}}^\prime\;,
%%%%%%%%%%%%%%%%%%%%%%%%%%%%%%%%%%%%%%%%%%%%%%%%%%%
\label{eff-Wilson-Coefficients}
\end{eqnarray}
where the concrete expression for $Y(q^2)$ can also be found in Ref. \cite{Buras1}.
In our numerical analyses, we evaluate the Wilson coefficients from the SM to
next-to-next-to-logarithmic (NNLL) accuracy in Table.\ref{tab1} at hadronic energy scale.
\begin{table}
\begin{tabular}{|c|c|c|c|}
\hline
\hline
$C_{_7}^{eff}$    & $C_{_8}^{eff}$    & $C_{_9}^{eff}-Y(q^2)$ & $C_{_{10}}^{eff}$\\
\hline
$-0.304$   & $-0.167$  & $4.211$ & $-4.103$\\
\hline
\hline
\end{tabular}
\caption{At hadronic scale $\mu=m_{_b}=4.8$GeV, SM Wilson coefficients to NNLL accuracy. \label{tab1}}
\end{table}
On the other hand, the corrections to the Wilson coefficients from new physics
are only included to one-loop accuracy:
\begin{eqnarray}
%%%%%%%%%%%%%%%%%%%%%%%%%%%%%%%%%%%%%%%%%%%%%%%%%%%
&&\overrightarrow{C}_{_{NP}}(\mu)=\widehat{U}(\mu,\mu_0)\overrightarrow{C}_{_{NP}}(\mu_0)
\;,\nonumber\\
%%%%%%%%%%%%%%%%%%%%%%%%%%%%%%%%%%%%%%%%%%%%%%%%%%%
&&\overrightarrow{C^\prime}_{_{NP}}(\mu)=\widehat{U^\prime}(\mu,\mu_0)
\overrightarrow{C^\prime}_{_{NP}}(\mu_0)
%%%%%%%%%%%%%%%%%%%%%%%%%%%%%%%%%%%%%%%%%%%%%%%%%%%
\label{evaluation1}
\end{eqnarray}
with
\begin{eqnarray}
%%%%%%%%%%%%%%%%%%%%%%%%%%%%%%%%%%%%%%%%%%%%%%%%%%%
&&\overrightarrow{C}_{_{NP}}^{T}=\Big(C_{_{1,NP}},\;\cdots,\;C_{_{6,NP}},
C_{_{7,NP}}^{eff},\;C_{_{8,NP}}^{eff},\;C_{_{9,NP}}^{eff}-Y_{_{NP}}(q^2),\;
C_{_{10,NP}}^{eff}\Big)
\;,\nonumber\\
%%%%%%%%%%%%%%%%%%%%%%%%%%%%%%%%%%%%%%%%%%%%%%%%%%%
&&\overrightarrow{C}_{_{NP}}^{\prime,\;T}=\Big(C_{_{7,NP}}^{\prime,\;eff},\;
C_{_{8,NP}}^{\prime,\;eff},\;C_{_{9,NP}}^{\prime,\;eff},\;
C_{_{10,NP}}^{\prime,\;eff}\Big)\;.
%%%%%%%%%%%%%%%%%%%%%%%%%%%%%%%%%%%%%%%%%%%%%%%%%%%
\label{evaluation2}
\end{eqnarray}
Correspondingly the evolving matrices are approached as
\begin{eqnarray}
%%%%%%%%%%%%%%%%%%%%%%%%%%%%%%%%%%%%%%%%%%%%%%%%%%%
&&\widehat{U}(\mu,\mu_0)\simeq1-\Big[{1\over2\beta_0}\ln{\alpha_{_s}(\mu)\over
\alpha_{_s}(\mu_0)}\Big]\widehat{\gamma}^{(0)T}
\;,\nonumber\\
%%%%%%%%%%%%%%%%%%%%%%%%%%%%%%%%%%%%%%%%%%%%%%%%%%%
&&\widehat{U^\prime}(\mu,\mu_0)\simeq1-\Big[{1\over2\beta_0}\ln{\alpha_{_s}(\mu)\over
\alpha_{_s}(\mu_0)}\Big]\widehat{\gamma^\prime}^{(0)T}\;,
%%%%%%%%%%%%%%%%%%%%%%%%%%%%%%%%%%%%%%%%%%%%%%%%%%%
\label{evaluation3}
\end{eqnarray}
where the anomalous dimension matrices can be read from Ref.  \cite{Gambino1} as
\begin{eqnarray}
%%%%%%%%%%%%%%%%%%%%%%%%%%%%%%%%%%%%%%%%%%%%%%%%%%%
&&\widehat{\gamma}^{(0)}=\left(\begin{array}{cccccccccc}
-4&{8\over3}&0&-{2\over9}&0&0&-{208\over243}&{173\over162}&-{2272\over729}&0\\
12&0&0&{4\over3}&0&0&{416\over81}&{70\over27}&{1952\over243}&0\\
0&0&0&-{52\over3}&0&2&-{176\over81}&{14\over27}&-{6752\over243}&0\\
0&0&-{40\over9}&-{100\over9}&{4\over9}&{5\over6}&-{152\over243}&-{587\over162}&-{2192\over729}&0\\
0&0&0&-{256\over3}&0&20&-{6272\over81}&{6596\over27}&-{84032\over243}&0\\
0&0&-{256\over9}&{56\over9}&{40\over9}&-{2\over3}&{4624\over243}&{4772\over81}&-{37856\over729}&0\\
0&0&0&0&0&0&{32\over3}&0&0&0\\
0&0&0&0&0&0&-{32\over9}&{28\over3}&0&0\\
0&0&0&0&0&0&0&0&0&0\\
0&0&0&0&0&0&0&0&0&0\\
\end{array}\right)
\;,\nonumber\\
%%%%%%%%%%%%%%%%%%%%%%%%%%%%%%%%%%%%%%%%%%%%%%%%%%%
&&\widehat{\gamma^\prime}^{(0)}=\left(\begin{array}{cccc}
{32\over3}&0&0&0\\
-{32\over9}&{28\over3}&0&0\\
0&0&0&0\\0&0&0&0\\
\end{array}\right)\;.
%%%%%%%%%%%%%%%%%%%%%%%%%%%%%%%%%%%%%%%%%%%%%%%%%%%
\label{ADM1}
\end{eqnarray}
In addition, the operators ${\cal O}_{_{S,P}}^{(\prime)}$ do not mix with other operators
and their Wilson coefficients are given by the corresponding coefficients at matching
scale.

\section{The branching ratios of $B_{s,d}^0\rightarrow\bar{l}l$ at hadronic scale
\label{sec4}}
\indent\indent

In the effective Hamilton Eq.~(\ref{effective-Hamilton}), the rare decays
$B_q^0\rightarrow\bar{l}l\:(l=\mu,\tau \:\textrm{and}\:q=s,\,d)$ are induced by the operators ${\cal O}_{_{9,10}}
,\;{\cal O}_{_{S,P}},\;{\cal O}_{_{9,10}}^\prime,\;{\cal O}_{_{S,P}}^\prime$
at hadronic scale. Correspondingly the hadronic matrix elements of axial vector
and pseudoscalar currents are parametrized as \cite{Dedes}
\begin{eqnarray}
%%%%%%%%%%%%%%%%%%%%%%%%%%%%%%%%%%%%%%%%%%%%%%%%%%%
&&\langle0|\bar{q}\gamma_\mu\gamma_5 b|B_q^0(p)\rangle=
ip^\mu f_{_{B_q^0}}
\;,\nonumber\\
%%%%%%%%%%%%%%%%%%%%%%%%%%%%%%%%%%%%%%%%%%%%%%%%%%%
&&\langle 0|\bar{q}0\gamma_5b|B_q^0(p)\rangle=
-i{M_{_{B_q^0}}^2f_{_{B_q^0}}\over m_{_b}+m_{_q}}\;,
%%%%%%%%%%%%%%%%%%%%%%%%%%%%%%%%%%%%%%%%%%%%%%%%%%%
\label{Hadronic-Matrix-Elements}
\end{eqnarray}
where $f_{_{B_q^0}}$ denote the decay constants respectively:
\begin{eqnarray}
%%%%%%%%%%%%%%%%%%%%%%%%%%%%%%%%%%%%%%%%%%%%%%%%%%%
f_{_{B_s^0}}=(227\pm8)\;{\rm MeV}
\;,\qquad f_{_{B_d^0}}=(190\pm8)\;{\rm MeV}\;,
%%%%%%%%%%%%%%%%%%%%%%%%%%%%%%%%%%%%%%%%%%%%%%%%%%%
\label{decay-constants}
\end{eqnarray}
and $M_{_{B_q^0}}$ denote the masses of neutral mesons
\begin{eqnarray}
%%%%%%%%%%%%%%%%%%%%%%%%%%%%%%%%%%%%%%%%%%%%%%%%%%%
M_{_{B_s^0}}=5.36677\;{\rm GeV}
\;,\qquad M_{_{B_d^0}}=5.27958\;{\rm GeV}\;.
%%%%%%%%%%%%%%%%%%%%%%%%%%%%%%%%%%%%%%%%%%%%%%%%%%%
\label{meson-masses}
\end{eqnarray}
Generally the matrix element ${\cal M}$ is expressed as:
\begin{eqnarray}
%%%%%%%%%%%%%%%%%%%%%%%%%%%%%%%%%%%%%%%%%%%%%%%%%%%
&&{\cal M}_{_q}={i4G_{_F}\over\sqrt{2}}V_{_{tb}}V_{_{ts}}^*\Big\{
F_{_S}^q\bar{l}l+F_{_P}^q\bar{l}\gamma_5l+F_{_V}^qp_\mu\bar{l}\gamma^\mu l
+F_{_A}^qp_\mu\bar{l}\gamma^\mu\gamma_5 l\Big\}\;,
%%%%%%%%%%%%%%%%%%%%%%%%%%%%%%%%%%%%%%%%%%%%%%%%%%%
\label{matrix-elements}
\end{eqnarray}
where the form factors $F_{_S}^q,\;F_{_P}^q,\;F_{_V}^q,\;F_{_A}^q$ of the scalar,
pseudoscalar, vector and axial-vector currents are given
\begin{eqnarray}
%%%%%%%%%%%%%%%%%%%%%%%%%%%%%%%%%%%%%%%%%%%%%%%%%%%
&&F_{_S}^q={\alpha_{_{EW}}(\mu_{_b})\over8\pi}{m_{_b}M_{_{B_q^0}}^2\over m_{_b}+m_{_q}}f_{_{B_q^0}}
(C_{_S}-C_{_S}^\prime)
\;,\nonumber\\
%%%%%%%%%%%%%%%%%%%%%%%%%%%%%%%%%%%%%%%%%%%%%%%%%%%
&&F_{_P}^q={\alpha_{_{EW}}(\mu_{_b})\over8\pi}{m_{_b}M_{_{B_q^0}}^2\over m_{_b}+m_{_q}}f_{_{B_q^0}}
(C_{_P}-C_{_P}^\prime)
\;,\nonumber\\
%%%%%%%%%%%%%%%%%%%%%%%%%%%%%%%%%%%%%%%%%%%%%%%%%%%
&&F_{_V}^q={\alpha_{_{EW}}(\mu_{_b})\over8\pi}f_{_{B_q^0}}
\Big[C_{_9}^{eff}(\mu_{_b})-C_{_9}^{\prime eff}(\mu_{_b})\Big]
\;,\nonumber\\
%%%%%%%%%%%%%%%%%%%%%%%%%%%%%%%%%%%%%%%%%%%%%%%%%%%
&&F_{_A}^q={\alpha_{_{EW}}(\mu_{_b})\over8\pi}f_{_{B_q^0}}
\Big[C_{_{10}}^{eff}(\mu_{_b})-C_{_{10}}^{\prime eff}(\mu_{_b})\Big]\;.
%%%%%%%%%%%%%%%%%%%%%%%%%%%%%%%%%%%%%%%%%%%%%%%%%%%
\label{form-factors}
\end{eqnarray}
Correspondingly the squared amplitude is
\begin{eqnarray}
%%%%%%%%%%%%%%%%%%%%%%%%%%%%%%%%%%%%%%%%%%%%%%%%%%%
&&|{\cal M}_{_q}|^2=16G_{_F}^2|V_{_{tb}}V_{_{ts}}^*|^2M_{_{B_q^0}}^2
\Big\{|F_{_S}^q|^2+|F_{_P}^q+2m_{_l}F_{_A}^q|^2\Big\}\;.
%%%%%%%%%%%%%%%%%%%%%%%%%%%%%%%%%%%%%%%%%%%%%%%%%%%
\label{squared-amplitudes}
\end{eqnarray}
The branching ratio is then given by
\begin{eqnarray}
%%%%%%%%%%%%%%%%%%%%%%%%%%%%%%%%%%%%%%%%%%%%%%%%%%%
&&BR(B_{_q}^0\rightarrow \bar{l}l)={\tau_{_{B_q^0}}\over16\pi}
{|{\cal M}_{_q}|^2\over M_{_{B_q^0}}}\sqrt{1-{4m_{_l}^2\over M_{_{B_q^0}}^2}}
%%%%%%%%%%%%%%%%%%%%%%%%%%%%%%%%%%%%%%%%%%%%%%%%%%%
\label{kappa-Factor}
\end{eqnarray}
with $\tau_{_{B_s^0}}=1.466(31)\;{\rm ps},\;\tau_{_{B_d^0}}=1.519(7)\;{\rm ps}$
denoting the life time of mesons.

In generic new physics, the branching ratio $BR(B_q^0\rightarrow\bar{l}l)_{_{NP}}$
is sensitive to the operators $C_{_{10}}^{(\prime)}$ and ${\cal O}^{(\prime)}_{_{S,P}}$:
\begin{eqnarray}
%%%%%%%%%%%%%%%%%%%%%%%%%%%%%%%%%%%%%%%%%%%%%%%%%%%
&&{BR(B_{_q}^0\rightarrow \bar{l}l)_{_{NP}}\over
BR(B_{_q}^0\rightarrow \bar{l}l)_{_{SM}}}=|S|^2\Big(1-{4m_{_l}^2\over M_{_{B_q^0}}^2}\Big)
+|P|^2
%%%%%%%%%%%%%%%%%%%%%%%%%%%%%%%%%%%%%%%%%%%%%%%%%%%
\label{NP-Bsmu+mu-}
\end{eqnarray}
with
\begin{eqnarray}
%%%%%%%%%%%%%%%%%%%%%%%%%%%%%%%%%%%%%%%%%%%%%%%%%%%
&&S\simeq{M_{_{B_q^0}}^2\over2m_{_l}}\cdot{C_{_{S}}-C_{_{S}}^\prime\over|C_{_{10,SM}}^{eff}(\mu_{_b})|}\;,
\nonumber\\
&&P\simeq{M_{_{B_q^0}}^2\over2m_{_l}}\cdot{C_{_{P}}-C_{_{P}}^\prime\over|C_{_{10,SM}}^{eff}(\mu_{_b})|}
+{C_{_{10}}^{eff}(\mu_{_b})-C_{_{10}}^{\prime eff}(\mu_{_b})\over |C_{_{10,SM}}^{eff}(\mu_{_b})|}\;.
%%%%%%%%%%%%%%%%%%%%%%%%%%%%%%%%%%%%%%%%%%%%%%%%%%%
\label{NP-Bsmu+mu-1}
\end{eqnarray}

%-----------------------------------------------------------------------------
\begin{table}[t]
\renewcommand{\arraystretch}{1.3}
\centering
\begin{tabular}{|c|c|}
\hline
Input & Input \\
\hline
$m_{_B}=5.280$ GeV & $m_{_{K^*}}=0.896$ GeV\\
$m_{_{B_s}}=5.367$ GeV & $m_{_\mu}=0.106$ GeV\\
$m_{_{\rm W}}=80.40$ GeV & $m_{_{\rm Z}}=91.19$ GeV\\
$\tau_{_B}=2.307\times10^{12}$ GeV & $f_{_B}=0.190\pm0.004$\\
$\alpha_{_{\rm S}}(m_{_{\rm Z}})=0.118\pm0.002$ & $\alpha_{_{\rm S}}(m_{_{\rm Z}})=1/128.9$\\
$m_c(m_c)=1.27\pm0.11$ GeV & $m_b(m_b)=4.18\pm0.17$ GeV\\
$m_t^{pole}=173.1\pm1.3$ GeV &  \\
$\lambda_{_{CKM}}=0.225\pm0.001$  & $A_{_{CKM}}=0.811\pm0.022$\\
$\bar{\rho}=0.131\pm0.026$  & $\bar{\eta}=0.345\pm0.014$\\

\hline
\end{tabular}
\caption{Input parameters \cite{PDG} used in the numerical analysis\label{tab2}}
\label{InputSM}
\end{table}
%-----------------------------------------------------------------------------

\section{Numerical analyses\label{sec5}}
\indent\indent
For the experimental observations in $\bar{B}\rightarrow X_{_s}\gamma$ and
$B_s^0\rightarrow l^+l^-$, the relevant SM inputs are presented in table.\ref{tab2}.
The supersymmetric parameters involved here are soft breaking masses
of the 2nd and 3rd generation squarks, $m_{_{\tilde{Q}_{2,3}}}^2,\;m_{_{\tilde{U}_{2,3}}}^2,\;
m_{_{\tilde{D}_{2,3}}}^2$, neutralino and chargino masses $m_{_{\chi_\alpha^0}},\;
m_{_{\chi_\beta^\pm}},\;(\alpha=1,\;\cdots,\;4,\;\beta=1,\;2)$ and their mixing matrices.
Additionally the free parameters also include $B-L$ gaugino/right-handed neutrino masses
and mixing which are mainly determined
from the nonzero VEVs of right-handed sneutrinos, the local $B-L$ gauge coupling $g_{_{BL}}$
and the soft gaugino mass $m_{_{BL}}$. The flavor conservation mixing between left- and right-handed
squarks $(\delta_u^{LR})_{33}=m_{_{\tilde{t}_X}}^2/\Lambda_{_{NP}}^2,\;
(\delta_d^{LR})_{33}=m_{_{\tilde{b}_X}}^2/\Lambda_{_{NP}}^2$ are chosen to give
the lightest Higgs mass in the range 124--126 GeV, where $\Lambda_{_{NP}}$ represents
the energy scale of supersymmetry and the concrete expressions
of $m_{_{\tilde{t}_X}}^2,\;m_{_{\tilde{b}_X}}^2$ are presented in appendix \ref{appA}.
The $b\rightarrow s$ transitions are mediated by those flavor changing insertions
$(\delta_{U,D}^{LL})_{23}=(\delta m_{_{\tilde{U},\tilde{D}}}^2)_{_{23}}^{LL}/\Lambda_{_{NP}}^2$,
$(\delta_{U,D}^{LR})_{23}=(\delta m_{_{\tilde{U},\tilde{D}}}^2)_{_{23}}^{LR}/\Lambda_{_{NP}}^2$,
$(\delta_{U,D}^{RR})_{23}=(\delta m_{_{\tilde{U},\tilde{D}}}^2)_{_{23}}^{RR}/\Lambda_{_{NP}}^2$,
which are originated from flavour-violating scalar mass terms and trilinear scalar couplings
in soft breaking terms.

To coincide with updated experimental data on supersymmetric particle searching from LHC
etc. \cite{PDG}, we choose $m_{_{\tilde{Q}_{2}}}=m_{_{\tilde{Q}_{3}}}=m_{_{\tilde{U}_{2}}}
=m_{_{\tilde{D}_{2}}}=m_{_{\tilde{D}_{3}}}=2\;{\rm TeV}$, $m_{_{\tilde{U}_{3}}}=1\;{\rm TeV}$,
$\Lambda_{_{NP}}=A_{_\tau}=A_{_b}=1\;{\rm TeV}$. For those parameters in Higgsino and gaugino sectors of the MSSM, we set
$m_1=200\;{\rm GeV}$, $m_2=400\;{\rm GeV}$, $m_{_{\tilde g}}=2\;{\rm TeV}$, $\mu=600\;{\rm GeV}$.
For the gauge coupling of local $B-L$ symmetry and relevant gaugino mass,
we take $g_{_{BL}}=0.7$, $m_{_{BL}}=0.5\;{\rm TeV}$, $\upsilon_{_N}=(0,\;0,\;3)\;{\rm TeV}$ here.
Similar to scenarios of the MSSM, the $b\rightarrow s\gamma$ transition can be evoked by the insertions
$(\delta_{U}^{LL})_{23},\;(\delta_{U}^{LR})_{23},\;(\delta_{U}^{RR})_{23}$ through one loop
diagrams composed by virtual charginos and up-type scalar quarks,
which are extensively discussed in literature before. In order to simplify our
analyses here, we choose $(\delta_{U}^{LL})_{23}=(\delta_{U}^{RR})_{23}=(\delta_{U}^{LR})_{23}=0$
unless a particular specification being made. Actually the numerical results of
$BR(B_s^0\rightarrow l^+l^-),\;(l=\mu,\;\tau)$ depend on the insertion
$(\delta_{D}^{LR})_{23}$ and CP phase $\theta_{_{BL}}$ mildly with this choice on
the parameter space. Because of the reason above, we set $(\delta_{D}^{LR})_{23}=\theta_{_{BL}}=0$
and mass of the lightest CP-odd Higgs
as $m_{_{A^0}}=1\;{\rm TeV}$. With those assumptions on parameters of the model
considered here, one obtains theoretical prediction on the lightest CP-even Higgs mass
around the value $125$ GeV as $\tan\beta=40$ partnering with $A_{_t}=0.5\;{\rm TeV}$,
$\tan\beta=20$ partnering with $A_{_t}=0.6\;{\rm TeV}$, or $\tan\beta=10$
partnering with $A_{_t}=1\;{\rm TeV}$ respectively,
which coincides with the experimental data from LHC.

It is well known that the experimental observation on
$BR(\bar{B}\rightarrow X_{_s}\gamma)$ constrains the relevant parameters strongly,
the average experimental data on the branching ratio of the inclusive
$\bar{B}\rightarrow X_{_s}\gamma$ reads  \cite{PDG}
\begin{eqnarray}
%%%%%%%%%%%%%%%%%%%%%%%%%%%%%%%%%%%%%%%%%%%%%%%%%%%
&&BR(\bar{B}\rightarrow X_{_s}\gamma)_{_{EXP}}=(3.40\pm0.21)\times10^{-4}\;,
%%%%%%%%%%%%%%%%%%%%%%%%%%%%%%%%%%%%%%%%%%%%%%%%%%%
\label{EXP-BtoSGamma}
\end{eqnarray}
which is consistent with the correspondingly SM prediction at NNLO order  \cite{Misiak1,Misiak2}
\begin{eqnarray}
%%%%%%%%%%%%%%%%%%%%%%%%%%%%%%%%%%%%%%%%%%%%%%%%%%%
&&BR(\bar{B}\rightarrow X_{_s}\gamma)_{_{SM}}=(3.36\pm0.23)\times10^{-4}\;.
%%%%%%%%%%%%%%%%%%%%%%%%%%%%%%%%%%%%%%%%%%%%%%%%%%%
\label{SM-BtoSGamma}
\end{eqnarray}
Through scanning the parameter space, we find that theoretical predictions
on the branching ratio of $\bar{B}\rightarrow X_{_s}\gamma$
depends on the insertions $(\delta_{D}^{LL})_{23},\;(\delta_{D}^{RR})_{23}$ weakly
in the model considered here.

%%%%%%%%%%%%%%%%%%%%%%%%%%%%%%%%%%%%%%%%%%%%%%%%%%%%%
\begin{figure}
\setlength{\unitlength}{1mm}
\centering
\vspace{0.0cm}
\includegraphics[width=3.2in]{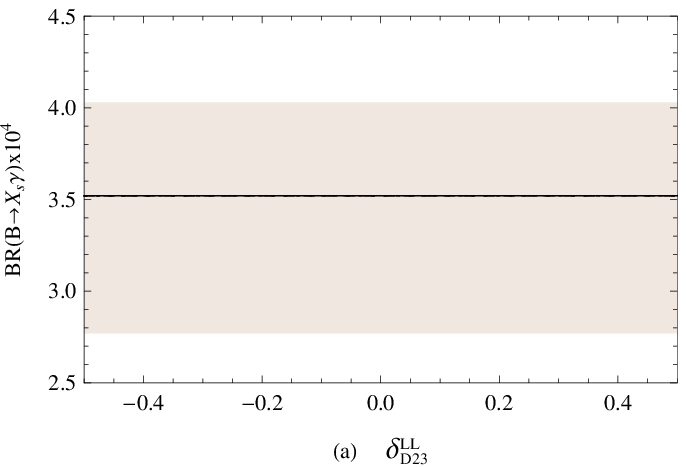}
\vspace{0.5cm}
\includegraphics[width=3.2in]{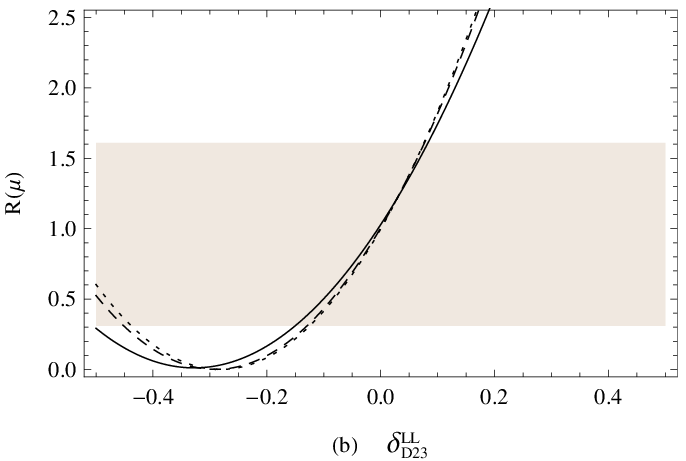}
\vspace{0.5cm}
\includegraphics[width=3.2in]{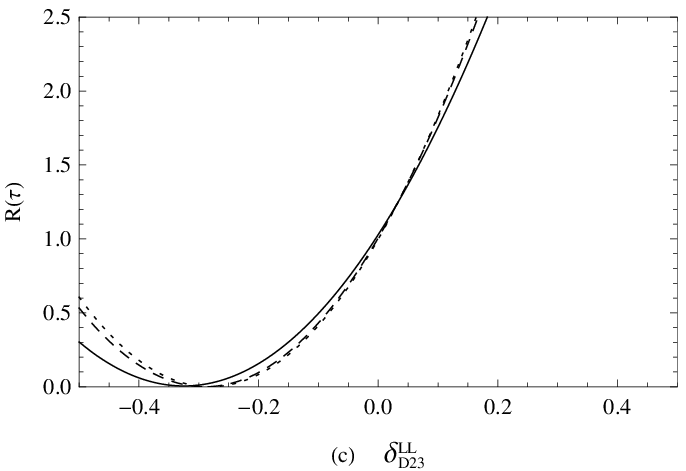}
\vspace{0cm}
\caption[]{Taking $\theta_{_9}=0,\;(\delta_{D}^{RR})_{23}=0$,
we plot $BR(\bar{B}\rightarrow X_{_s}\gamma)$, $R(\mu)$ and $R(\tau)$
varying with the insertion $(\delta_{D}^{LL})_{23}$
in (a), (b) and (c), respectively. Where the solid line represents
$\tan\beta=40,\;A_{_t}=0.5\;{\rm TeV}$, the dashed line represents
$\tan\beta=20,\;A_{_t}=0.6\;{\rm TeV}$, and the dotted line represents
$\tan\beta=10,\;A_{_t}=1\;{\rm TeV}$. In addition, the gray regions
represent the experimental results within $3\sigma$ permission.}
\label{fig1}
\end{figure}
%%%%%%%%%%%%%%%%%%%%%%%%%%%%%%%%%%%%%%%%%%%%%%%%%%%%%

%%%%%%%%%%%%%%%%%%%%%%%%%%%%%%%%%%%%%%%%%%%%%%%%%%%%%
\begin{figure}
\setlength{\unitlength}{1mm}
\centering
\vspace{0.0cm}
\includegraphics[width=3.2in]{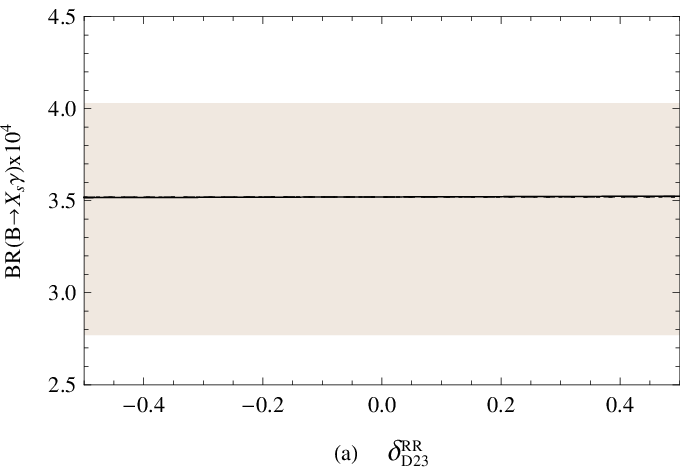}
\vspace{0.5cm}
\includegraphics[width=3.2in]{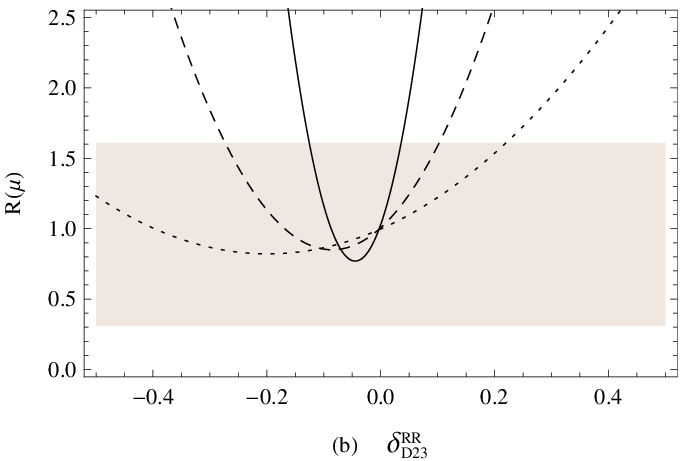}
\vspace{0.5cm}
\includegraphics[width=3.2in]{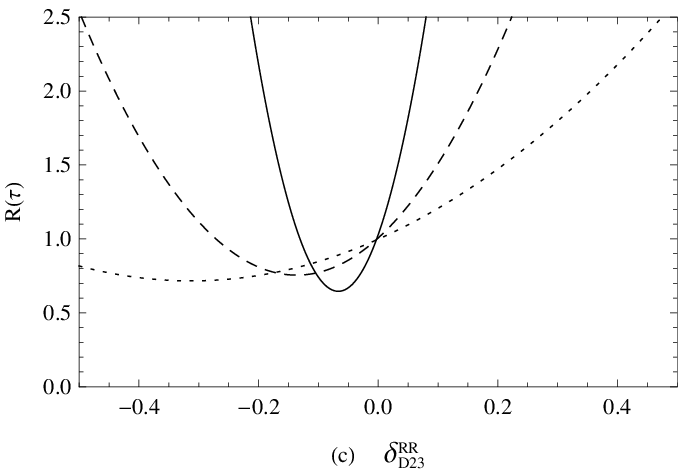}
\vspace{0cm}
\caption[]{Taking  $\theta_{_9}=0,\;(\delta_{D}^{LL})_{23}=0$,
we plot $BR(\bar{B}\rightarrow X_{_s}\gamma)$, $R(\mu)$ and $R(\tau)$
varying with the insertion $(\delta_{D}^{RR})_{23}$
in (a), (b) and (c), respectively. Where the solid line represents
$\tan\beta=40,\;A_{_t}=0.5\;{\rm TeV}$, the dashed line represents
$\tan\beta=20,\;A_{_t}=0.6\;{\rm TeV}$, and the dotted line represents
$\tan\beta=10,\;A_{_t}=1\;{\rm TeV}$. In addition, the gray regions
represent the experimental results within $3\sigma$ permission.}
\label{fig2}
\end{figure}
%%%%%%%%%%%%%%%%%%%%%%%%%%%%%%%%%%%%%%%%%%%%%%%%%%%%%

%%%%%%%%%%%%%%%%%%%%%%%%%%%%%%%%%%%%%%%%%%%%%%%%%%%%%
\begin{figure}
\setlength{\unitlength}{1mm}
\centering
\vspace{0.0cm}
\includegraphics[width=3.2in]{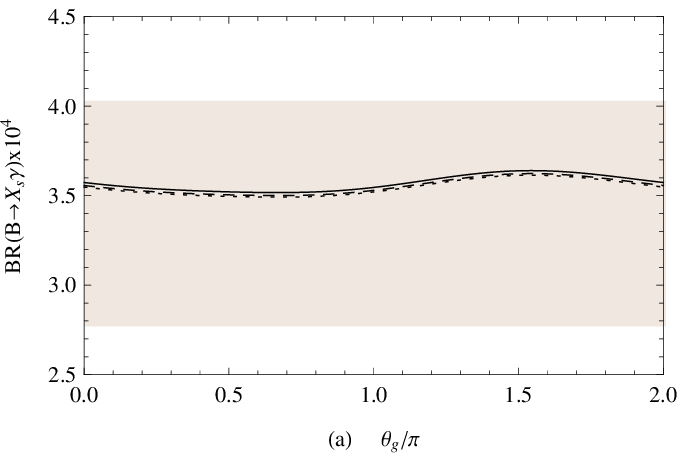}
\vspace{0.5cm}
\includegraphics[width=3.2in]{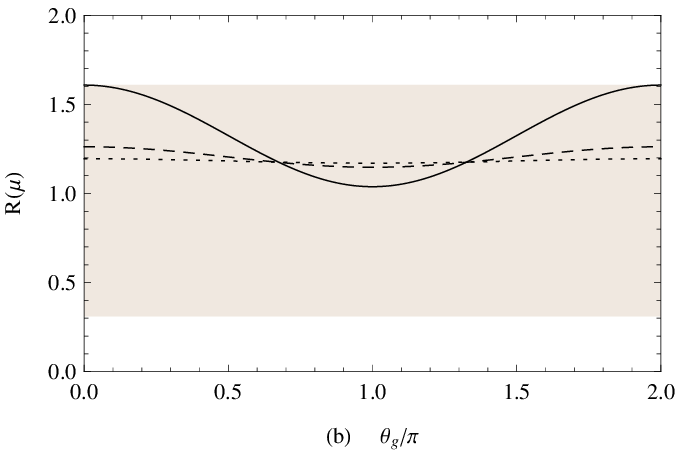}
\vspace{0.5cm}
\includegraphics[width=3.2in]{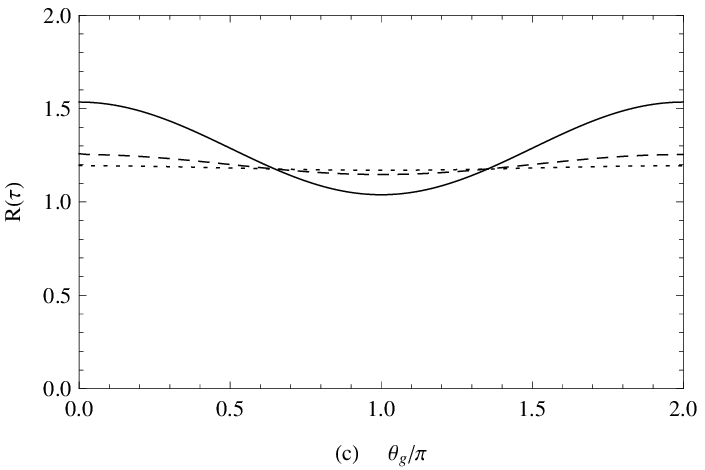}
\vspace{0cm}
\caption[]{Taking $(\delta_{U}^{LL})_{23}=(\delta_{U}^{RR})_{23}=(\delta_{D}^{LL})_{23}=(\delta_{D}^{RR})_{23}=0.02$ and
$(\delta_{U}^{LR})_{23}=(\delta_{D}^{LR})_{23}=-0.04$, we plot
$BR(\bar{B}\rightarrow X_{_s}\gamma)$, $R(\mu)$ and $R(\tau)$
varying with the CP phase $\theta_{_g}$
in (a), (b) and (c), respectively. Where the solid line represents
$\tan\beta=40,\;A_{_t}=0.5\;{\rm TeV}$, the dashed line represents
$\tan\beta=20,\;A_{_t}=0.6\;{\rm TeV}$, and the dotted line represents
$\tan\beta=10,\;A_{_t}=1\;{\rm TeV}$. In addition, the gray regions
represent the experimental results within $3\sigma$ permission.}
\label{fig3}
\end{figure}
%%%%%%%%%%%%%%%%%%%%%%%%%%%%%%%%%%%%%%%%%%%%%%%%%%%%%

Under our assumptions on the relevant parameter space, the supersymmetric corrections
to the $b\rightarrow sl^+l^-$ transition are mainly originated from
the insertions $(\delta_{D}^{LL})_{23},\;(\delta_{D}^{RR})_{23}$ through one loop
diagrams composed by $U(1)_{_{B-L}}$ gaugino/gluino and down type squarks of the 2nd and 3rd
generations.
Assuming CP phase $\theta_{_9}=0,\;(\delta_{D}^{RR})_{23}=0$,
we plot $BR(\bar{B}\rightarrow X_{_s}\gamma)$,
$R(\mu)=BR(B_{_s}^0\rightarrow \mu^-\mu^+)_{_{NP}}/
BR(B_{_s}^0\rightarrow \mu^-\mu^+)_{_{SM}}$ and $R(\tau)=BR(B_{_s}^0\rightarrow \tau^-\tau^+)_{_{NP}}/
BR(B_{_s}^0\rightarrow \tau^-\tau^+)_{_{SM}}$ varying with
$(\delta_{D}^{LL})_{23}$ in Fig.~\ref{fig1}, where the gray regions denote the experimental data
within 3 standard deviations. The new physics corrections to $BR(\bar{B}\rightarrow X_{_s}\gamma)$
mainly originate from the insertion $(\delta_{U}^{LR})_{23}$ through Feynman
diagrams composed by virtual chargino-stop particles, and theoretical evaluations
for $BR(\bar{B}\rightarrow X_{_s}\gamma)$ depends on $(\delta_{D}^{LL})_{23}$ mildly.
Meanwhile the experimental data on $BR(B_{_s}^0\rightarrow \mu^-\mu^+)$
favor the insertion $(\delta_{D}^{LL})_{23}$ lying in the range
$-0.1\le(\delta_{D}^{LL})_{23}\le0.1$. In limit of large $\tan\beta$,
dominating corrections to the Wilson coefficients $C_{_{S,P}}^{(\prime)}$
from new physics are proportional to the mass of lepton in final states $m_{_l}$.
Nevertheless the dependence on $m_{_l}$ is compensated by $m_{_l}$
from denominator in the first terms of $S,\;P$ respectively in Eq.~(\ref{NP-Bsmu+mu-1}).
Because of the reason, the theoretical evaluations on $R(\tau)$ are
not differ from that on $R(\mu)$ obviously.

In Fig.~\ref{fig2},
we plot $BR(\bar{B}\rightarrow X_{_s}\gamma)$, $R(\mu)$ and $R(\tau)$ varying with
$(\delta_{D}^{RR})_{23}$, as $\theta_{_9}=0$, $(\delta_{D}^{LL})_{23}=0$.
The theoretical evaluations
of $BR(\bar{B}\rightarrow X_{_s}\gamma)$ depends on $(\delta_{D}^{RR})_{23}$ mildly, too.
Meanwhile the experimental data on $BR(B_{_s}^0\rightarrow \mu^-\mu^+)$
favor the insertion $(\delta_{D}^{RR})_{23}$ lying in the ranges
$-0.1\le(\delta_{D}^{RR})_{23}\le0.05$ as $\tan\beta=40,\;A_{_t}=0.5\;{\rm TeV}$,
$-0.25\le(\delta_{D}^{RR})_{23}\le0.1$ as $\tan\beta=20,\;A_{_t}=0.6\;{\rm TeV}$,
and $-0.6\le(\delta_{D}^{RR})_{23}\le0.2$ as $\tan\beta=10,\;A_{_t}=1\;{\rm TeV}$, respectively.
Because of the reason mentioned above, the theoretical evaluations on $R(\tau)$ are
not differ from that on $R(\mu)$ obviously.

Taking $(\delta_{U}^{LL})_{23}=(\delta_{U}^{RR})_{23}=(\delta_{D}^{LL})_{23}=(\delta_{D}^{RR})_{23}=0.02$ and
$(\delta_{U}^{LR})_{23}=(\delta_{D}^{LR})_{23}=-0.04$,
we plot $BR(\bar{B}\rightarrow X_{_s}\gamma)$, $R(\mu)$ and $R(\tau)$ varying with
the CP phase $\theta_{_g}$ in Fig.~\ref{fig3}.
In Fig.~\ref{fig3}(a,b) the gray regions
represents the experimental data on $BR(\bar{B}\rightarrow X_{_s}\gamma)$
and $R(\mu)$ within 3 standard deviations, respectively. Adopting our assumptions on relevant parameter
space, one finds that those theoretical evaluations on $R(\mu)$ and $R(\tau)$
depend on the CP phase $\theta_{_g}$ acutely as $\tan\beta=40$.
Along with decreasing of $\tan\beta$, those numerical evaluations on $R(\mu)$ and $R(\tau)$
vary with the CP phase $\theta_{_g}$ mildly.

\section{Summary\label{sec6}}
\indent\indent
Considering the constraint from the observed Higgs signal at the LHC,
we study the supersymmetric corrections to the branching ratios
$BR(\bar{B}\rightarrow X_{_s}\gamma)$, $BR(B_q^0\rightarrow l^+l^-)\;(l=\mu,\;\tau)$
in the MSSM with local $U(1)_{B-L}$ symmetry
with nonuniversal soft breaking terms.
Under our assumptions on parameters of the considered model,
the numerical analyses indicate that the insertions $(\delta_{D}^{LL})_{23},\;(\delta_{D}^{RR})_{23}$
affects the theoretical predictions on $BR(B_q^0\rightarrow l^+l^-)\;(l=\mu,\;\tau)$ strongly
when the numerical evaluations of $BR(\bar{B}\rightarrow X_{_s}\gamma)$
are coincide with corresponding experimental observations.
In addition, the CP phase $\theta_{_g}$ also affects
the numerical results acutely when the neutral gauginos
$m_{_{\tilde g}}\sim m_{_{BL}}\ge 1\;{\rm TeV}$ and the squarks
acquire the masses around several ${\rm TeV}s$ in large $\tan\beta$ scenarios.

\begin{acknowledgments}
\indent\indent
The work has been supported by the National Natural
Science Foundation of China (NNSFC) with Grants No. 11275036, No. 11535002,
the open project of State Key Laboratory of of Theoretical Physics,
Institute of Theoretical Physics, Chinese Academy of Sciences, China (No.Y5KF131CJ1),
the Natural Science Foundation of Hebei province with Grants No. A2013201277, No. A2016201069,
and the Found of Hebei province with the Grant NO. BR2-201 and
the Natural Science Fund of Hebei University with Grants No. 2011JQ05 and No. 2012-242,
Hebei Key Lab of Optic-Electronic Information and Materials, the midwest universities
comprehensive strength promotion project.
\end{acknowledgments}

\appendix

\section{The mass squared matrices for squarks\label{appA}}
\indent\indent
With the minimal flavor violation assumption, the $2\times2$ mass squared matrix
for scalar tops is given as
\begin{eqnarray}
&&{\cal Z}_{_t}^\dagger\left(\begin{array}{cc}
m_{_{\tilde{t}_L}}^2\;\;&m_{_{\tilde{t}_X}}^2\\ \\
m_{_{\tilde{t}_X}}^2\;\;&m_{_{\tilde{t}_R}}^2
\end{array}\right){\cal Z}_{_t}=diag\Big(m_{_{\tilde{t}_1}}^2,\;m_{_{\tilde{t}_2}}^2\Big)\;,
\label{UP-squark1}
\end{eqnarray}
with
\begin{eqnarray}
&&m_{_{\tilde{t}_L}}^2={(g_1^2+g_2^2)\upsilon_{_{\rm EW}}^2\over24}
\Big(1-2\cos^2\beta\Big)\Big(1-4c_{_{\rm W}}^2\Big)
\nonumber\\
&&\hspace{1.4cm}
+{g_{_{BL}}^2\over6}\Big(\upsilon_{_N}^2-\upsilon_{_{\rm EW}}^2
+\upsilon_{_{\rm SM}}^2\Big)+m_{_t}^2+m_{_{\tilde{Q}_3}}^2
\;,\nonumber\\
&&m_{_{\tilde{t}_R}}^2=-{g_1^2\upsilon_{_{\rm EW}}^2\over6}
\Big(1-2\cos^2\beta\Big)
\nonumber\\
&&\hspace{1.4cm}
-{g_{_{BL}}^2\over6}\Big(\upsilon_{_N}^2-\upsilon_{_{\rm EW}}^2
+\upsilon_{_{\rm SM}}^2\Big)
+m_{_t}^2+m_{_{\tilde{U}_3}}^2\;,
\nonumber\\
&&m_{_{\tilde{t}_X}}^2=-{\upsilon_{_u}\over\sqrt{2}}A_{_t}Y_{_t}
+{\mu\upsilon_{_d}\over\sqrt{2}}Y_{_t}\;.
\label{UP-squark2}
\end{eqnarray}
Here $Y_{_t},\;A_{_t}$ denote Yukawa coupling
and trilinear soft-breaking parameters in top quark sector, respectively.
In a similar way, the mass-squared matrix for scalar bottoms is
\begin{eqnarray}
&&{\cal Z}_{_b}^\dagger\left(\begin{array}{cc}
m_{_{\tilde{b}_L}}^2\;\;&m_{_{\tilde{b}_X}}^2\\ \\
m_{_{\tilde{b}_X}}^2\;\;&m_{_{\tilde{b}_R}}^2
\end{array}\right){\cal Z}_{_b}=diag\Big(m_{_{\tilde{b}_1}}^2,\;m_{_{\tilde{b}_2}}^2\Big)\;,
\label{DOWN-squark1}
\end{eqnarray}
with
\begin{eqnarray}
&&m_{_{\tilde{b}_L}}^2={(g_1^2+g_2^2)\upsilon_{_{\rm EW}}^2\over24}
\Big(1-2\cos^2\beta\Big)\Big(1+2c_{_{\rm W}}^2\Big)
\nonumber\\
&&\hspace{1.4cm}
+{g_{_{BL}}^2\over6}\Big(\upsilon_{_N}^2-\upsilon_{_{\rm EW}}^2
+\upsilon_{_{\rm SM}}^2\Big)
+m_{_b}^2+m_{_{\tilde{Q}_3}}^2\;,
\nonumber\\
&&m_{_{\tilde{b}_R}}^2={g_1^2\upsilon_{_{\rm EW}}^2\over12}
\Big(1-2\cos^2\beta\Big)
\nonumber\\
&&\hspace{1.4cm}
-{g_{_{BL}}^2\over6}\Big(\upsilon_{_N}^2-\upsilon_{_{\rm EW}}^2
+\upsilon_{_{\rm SM}}^2\Big)
+m_{_b}^2+m_{_{\tilde{D}_3}}^2\;,
\nonumber\\
&&m_{_{\tilde{b}_X}}^2={\upsilon_{_d}\over\sqrt{2}}A_{_b}Y_{_b}
-{\mu\upsilon_{_u}\over\sqrt{2}}Y_{_b}\;,
\label{DOWN-squark2}
\end{eqnarray}
here $Y_{_b},\;A_{_b}$ denote Yukawa couplings
and trilinear soft-breaking parameters in b quark sector, respectively.


\begin{thebibliography}{99}

\bibitem{LHCb-Bsmuons}R.~Aaij {\it et al.} (LHCb collaboration), Phys.~Rev.~Lett.~{\bf110}(2013)021801.
\bibitem{PDG}K.~A.~Olive {\it et al}. (Particle Data Group), Chin.~Phys.~C{\bf38}(2014)090001.
\bibitem{Buras0}A.~J.~Buras, J.~Girrbach, G.~Isidori, Eur.~Phys.~J.~C{\bf72}(2012)2172.
\bibitem{Gabrielli}E.~Gabrielli {\it et al.}, Phys.~Lett.~B{\bf374}(1996)80.
\bibitem{Ciuchini1}M.~Ciuchini, G.~Degrassi, P.~Gambino and G.~F.~Giudice, Nucl.~Phys.~B{\bf527}(1998)21.
\bibitem{Ciafaloni}P.~Ciafaloni, A.~Romanino and A.~Strumia, Nucl.~Phys.~B{\bf524}(1998)361.
\bibitem{Borzumati1}F.~Borzumati and C.~Greub, Phys.~Rev.~D{\bf58}(1998)074004.
\bibitem{Bertolini1}S.~Bertolini, F.~Borzumati, A.~Masiero and G.~Ridolfi, Nucl.~Phys.~B{\bf353}(1991)591.
\bibitem{Barbieri}R.~Barbieri and G.~F.~Giudice, Phys.~Lett.~B{\bf309}(1993)86.
\bibitem{Borzumati2}F.~Borzumati and C.~Greub, T.~Hurth and D.~Wyler, Phys.~Rev.~D{\bf62}(2000)075005.
\bibitem{Causse}M.~Causse and J.~Orloff, Eur.~Phys.~J.~C{\bf23}(2002)749.
\bibitem{Prelovsek}S.~Prelovsek and D.~Wyler, Phys.~Lett.~B{\bf500}(2001)304.
\bibitem{Ciuchini2}M.~Ciuchini, G.~Degrassi, P.~Gambino and G.~F.~Giudice, Nucl.~Phys.~B{\bf534}(1998)3.
\bibitem{Bertolini2}S.~Bertolini and J.~Matias, Phys.~Rev.~D{\bf57}(1998)4197.
\bibitem{Cottingham}W.~N.~Cottingham, H.~Mehrban and I.~B.~Whittingham, Phys.~Rev.~D{\bf60}(1999)114029.
\bibitem{Barenboim}G.~Barenboim and M.~Raidal, Phys.~Lett.~B{\bf457}(1999)109.
\bibitem{Hewett}J.~L.~Hewett and D.~Wells, Phys.~Rev.~D{\bf55}(1997)5549.
\bibitem{Ali}A.~Ali, P.~Ball, L.~T.~Handoko and G.~Hiller, Phys.~Rev.~D{\bf61}(2000)074024.
\bibitem{Buras1}W.~Altmannshofer, P.~Ball, A.~Bharucha, A.~J.~Buras, D.~M.~Straub, and M.~Wick,
JHEP {\bf 0901} (2009) 019, arXiv:0811.1214.
\bibitem{Masiero1}A.~Masiero and L.~Silvestrini, {\it Honolulu 1997, B physics and CP violation},
p.172, hep-ph/9709244.
\bibitem{Masiero1-1}A.~Masiero and L.~Silvestrini, {\it Erice 1997, Highlights of subnuclear physics}, p.404, hep-ph/9711401.
\bibitem{Ciuchini3}M.~Ciuchini {\it et al.}, JHEP {\bf9810} (1998) 008, arXiv:hep-ph/9808328.
\bibitem{Contion}R.~Contion, I.~Scimemi, Eur.~Phys.~J.~C{\bf10}(1999)347.
\bibitem{Krauss}F.~Krauss, G.~Soff,  Nucl.~Phys.~B{\bf633}(2002)237.
\bibitem{Feng1}Tai-Fu Feng, Xue-Qian Li, Wen-Gan Ma and Feng Zhang, Phys.~Rev.~D{\bf63}(2001)015013.
\bibitem{CMS} CMS Collaboration, Phys.~Lett.~B{\bf 716}(2012)30.
\bibitem{ATLAS}ATLAS Collaboration, Phys.~Lett.~B{\bf 716}(2012)1.
\bibitem{LHCB}B.~Adeva {\it et al.} (LHCb Collaboration), {\it Roadmap for selected key
measurements of LHCb},  arXiv:0912.4179.
\bibitem{B-factory1}T.~Aushev {\it et al.}, {\it Physics at Super B Factory},
arXiv:1002.5012.
\bibitem{B-factory2}B.~O'Leary {\it et al.} (SuperB Collaboration), {\it
SuperB Progress Reports}, arXiv:1008.1541.
\bibitem{R-parity}R.~Barbier {\it et.al.}, Phys.~Rep.~{\bf 420}(2005)1.
\bibitem{R-parity-1}C.-H. Chang, T.-F. Feng, Eur.~Phys.~J.~C{\bf 12}(2000)137.
\bibitem{Perez1}P.~Fileviez~Perez and S.~Spinner, Phys.~Lett.~B{\bf 673}(2009)251.
\bibitem{Perez2}V.~Barger, P.~Fileviez~Perez, and S.~Spinner, Phys.~Rev.~Lett.{\bf 102}(2009)181802.
\bibitem{Perez3}P.~Fileviez~Perez and S.~Spinner, Phys.~Rev.~D{\bf 80}(2009)015004.
\bibitem{Perez4}P.~Fileviez~Perez and S.~Spinner, JHEP {\bf 1204} (2012) 118, arXiv:1201.5923.
\bibitem{Perez6}V.~Barger, P.~Fileviez~Perez and S.~Spinner, Phys.~Lett.~B{\bf 696}(2011)509.
\bibitem{Senjanovic1}D.~K.~Ghosh, G.~Senjanovic, Y.~Zhang, Phys.~Lett.~B{\bf 698}(2011)420.
\bibitem{Chang-Feng}C.-H.~Chang, T.-F.~Feng, Y.-L.~Yan, H.-B.~Zhang, S.-M.~Zhao,
Phys.~Rev.~D{\bf90}(2014)035013.
\bibitem{Hamann}J.~Hamann, S.~Hannestad, G.~Raffelt, I.~Tamborra, and Y.~Y.~Y.~Wong,
Phys.~Rev.~Lett.{\bf 105}(2010)181301.
\bibitem{Feng2}T.-F.~Feng, Y.-L.~Yan, H.-B.~Zhang, S.-M.~Zhao,
Phys.~Rev.~D{\bf 92}(2015)055024.
\bibitem{Gambino1}P.~Gambino, M.~Gorbahn and U.~Haisch, Nucl.~Phys.~B{\bf 673}(2003)238.
\bibitem{Dedes}A.~Dedes, J.~Rosiek and P.~Tanedo, Phys.~Rev.~D{\bf79}(2009)055006.
\bibitem{Misiak1}M.~Misiak {\it et al.}, Phys.~Rev.~Lett.{\bf114}(2015)221801, arXiv:1503.01789.
\bibitem{Misiak2}M.~Czakon {\it et al.}, arXiv:1503.01791.

\end{thebibliography}
\end{document}